\documentclass[aps,prb,twocolumn,superscriptaddress]{revtex4-1}

\usepackage[usenames,dvipsnames]{xcolor}
\usepackage{subfig} 
\usepackage{placeins}

\usepackage{csquotes} 
\usepackage{upgreek} 
\usepackage[version=3]{mhchem} 
\usepackage[normalem]{ulem} 

\usepackage[pdftex,colorlinks,bookmarks = true]{hyperref} 
\usepackage[capitalize]{cleveref} 

\usepackage{amsmath}
\usepackage{amsfonts}
\usepackage{amssymb}
\usepackage{empheq} 
\usepackage{mathtools} 
\usepackage{bm}

\usepackage{siunitx} 
\usepackage{braket} 

\newcommand{\avg}[1]{\left\langle #1 \right\rangle} 
\let\vaccent=\v 
\renewcommand{\v}[1]{\bm{#1}}
\newcommand{\inv}{^{-1}} 
\newcommand{\e}[1]{\, \mathrm{e}^{#1}} 
\renewcommand{\i}{\mathrm{i}} 
\newcommand{\cpi}{\uppi} 
\providecommand*{\groupU}{\mathrm{U}(1)} 
\providecommand*{\groupZ}{\mathbb{Z}_2} 
\providecommand*{\groupUZ}{\groupU \times \groupZ} 

\makeatletter 
\providecommand*{\diff}
	  {\@ifnextchar^{\DIfF}{\DIfF^{}}}
  \def\DIfF^#1{%
	  \mathop{\mathrm{\mathstrut d}}%
		  \nolimits^{#1}\gobblespace}
  \def\gobblespace{%
	  \futurelet\diffarg\opspace}
  \def\opspace{%
	  \let\DiffSpace\!%
	  \ifx\diffarg(%
		  \let\DiffSpace\relax
	  \else
		  \ifx\diffarg[%
			  \let\DiffSpace\relax
		  \else
			  \ifx\diffarg\{%
				  \let\DiffSpace\relax
			  \fi\fi\fi\DiffSpace}
\makeatother

\begin{document}

\title{Josephson-frustrated superconductors in a magnetic field}

\author{Troels Arnfred Bojesen}
\email{troels.bojesen@ntnu.no}
\affiliation{Department of Physics, Norwegian University of Science and Technology, NO-7491 Trondheim, Norway}
\author{Asle Sudb\o{}}
\email{asle.sudbo@ntnu.no}
\affiliation{Department of Physics, Norwegian University of Science and Technology, NO-7491 Trondheim, Norway}

\date{\today}

\begin{abstract}
We study the effect of an externally imposed rotation or magnetic field on frustrated multiband superconductors/superfluids. The frustration originates with multiple superconducting bands crossing the Fermi surface in conjunction with interband Josephson-couplings with a positive sign. These couplings tend to frustrate the phases of the various components of the superconducting order parameter. This in turn leads to an effective description in terms of a $\groupUZ$-symmetric system, where essentially only the $\groupU$-sector couples to the gauge-field representing the rotation or magnetic field. By imposing a large enough net vorticity on the system at low temperatures, one may therefore reveal a resistive vortex liquid state which will feature an unusual additional phase transition in the $\groupZ$-sector. At low enough vorticity there is a corresponding vortex-lattice phase featuring a $\groupZ$ phase transition. We argue that this Ising transition phase should be readily observable in experiments.  
\end{abstract}

\pacs{}

\maketitle

\section{Introduction}
Multiband superconductors, that is superconductors with more than two superconducting bands crossing the Fermi-surface, \cite{seyfarth,hosono,Prakash,Lu,Liu,Kuroki,Subedi} may display fascinating physics which has no counterpart in single- or two-band superconductors, including the possibility of spontaneous breaking of time-reversal symmetry.\cite{zlatko, maiti} These phenomena originate with the interplay between phase-variables of each of the components of the superconducting order parameter: Having more than two fluctuating phase-degrees of freedom inherently leads to an internal frustration of the superconducting order parameter, provided the interband Josephson couplings are positive. Such phenomena are not seen in the single- or two-band cases.\cite{weston,PhysRevB.89.104509} Recently, it has been demonstrated that phase-frustrated systems feature phase diagrams which are a result of large fluctuations,\cite{bojesen_MF_CMF_MC} and as such are fundamentally not captured correctly by standard mean-field descriptions of these system, which ignore completely fluctuations in these phase-variables.

Phase-fluctuations come into play in a particularly important manner in Josephson-frustrated systems at least in two instances. The first case is close to thermally driven phase transitions in zero external field.\cite{PhysRevB.88.220511,PhysRevB.89.104509} The second is associated with the physics of field-induced topological defects of the superconducting order parameter components, which involve $2 \cpi$ phase-windings in the phase variables. In this paper, we will focus on the latter, and see how a tuning of the phase-transition in the lattice of field-induced topological defects (vortex lattice) of a multiband superconductor (or for that matter a multi-component superfluid or even a multi-component spinor Bose-Einstein condensate) may be used to unearth unexpected emergent broken symmetries in multiband superconductors. Prime examples of the multiband superconductors that we have in mind, are heavy fermion systems~\cite{seyfarth} and the more recently discovered iron-pnictide high-temperature superconductors,\cite{hosono,Prakash,Lu,Liu,Kuroki,Subedi} but our discussion will be applicable more generally to any system with a spinor-type order parameter with three or more components.  

When a container holding a (one component) superfluid liquid is subject to rotation, the circulation of the condensate is quantized into vortices parallel to the axis of rotation. These vortices may be described as externally imposed topological defects of the $\groupU$ order parameter field describing the condensate. This is in contrast to the thermally induced proliferation of vortex-antivortex pairs (2D) or vortex-loops (3D) driving the transition from a superfluid to a normal fluid. The vortices interact, and below a given temperature they will self-organize into a lattice structure. An equivalent situation is found in type II superconductors subject to an external magnetic field, where the topological defects form vortex lines of zeroes of the order parameter in addition to exhibiting tubes of confined and quantized magnetic flux. 

\begin{figure}[ht]
  \subfloat[Phases of the fields.\label{fig:phase_definition}]{
    \def\svgwidth{0.4\columnwidth}
  \begingroup%
  \makeatletter%
  \providecommand\color[2][]{%
    \errmessage{(Inkscape) Color is used for the text in Inkscape, but the package 'color.sty' is not loaded}%
    \renewcommand\color[2][]{}%
  }%
  \providecommand\transparent[1]{%
    \errmessage{(Inkscape) Transparency is used (non-zero) for the text in Inkscape, but the package 'transparent.sty' is not loaded}%
    \renewcommand\transparent[1]{}%
  }%
  \providecommand\rotatebox[2]{#2}%
  \ifx\svgwidth\undefined%
    \setlength{\unitlength}{135.16518555bp}%
    \ifx\svgscale\undefined%
      \relax%
    \else%
      \setlength{\unitlength}{\unitlength * \real{\svgscale}}%
    \fi%
  \else%
    \setlength{\unitlength}{\svgwidth}%
  \fi%
  \global\let\svgwidth\undefined%
  \global\let\svgscale\undefined%
  \makeatother%
  \begin{picture}(1,0.59557126)%
    \put(0,0){\includegraphics[width=\unitlength]{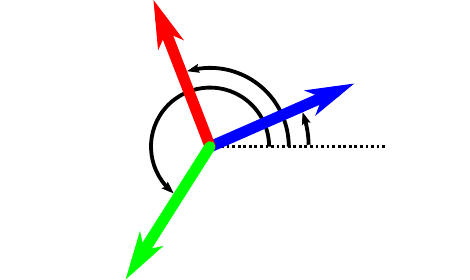}}%
    \put(0.68813623,0.30221329){\color[rgb]{0,0,0}\makebox(0,0)[lb]{\smash{$\theta_{1}$}}}%
    \put(0.5098767,0.4595013){\color[rgb]{0,0,0}\makebox(0,0)[lb]{\smash{$\theta_{2}$}}}%
    \put(0.31188689,0.31269923){\color[rgb]{0,0,0}\makebox(0,0)[rb]{\smash{$\theta_{3}$}}}%
  \end{picture}%
  \endgroup%
  }\\
  \subfloat[$+1$\label{fig:3C_ground_state_1}]{\includegraphics[width=0.4\columnwidth]{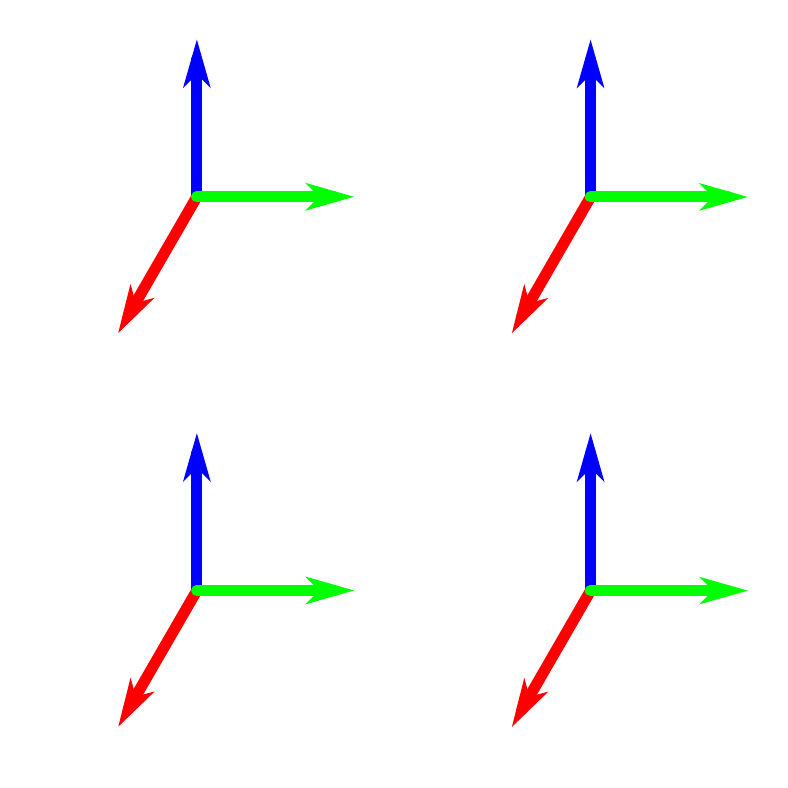}}
  \qquad
  \subfloat[$-1$\label{fig:3C_ground_state_2}]{\includegraphics[width=0.4\columnwidth]{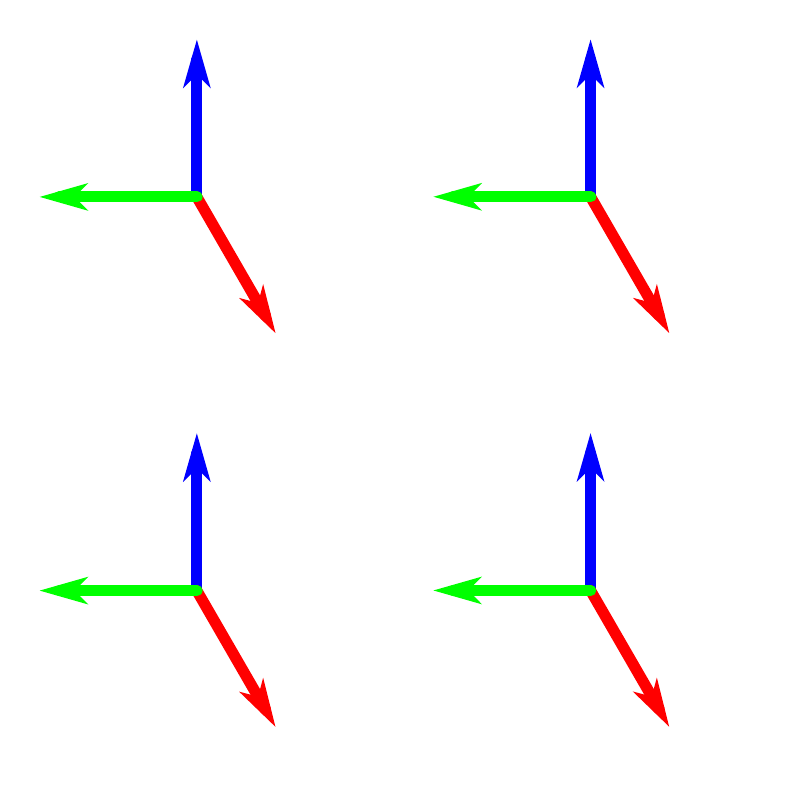}}
  \caption{(Colors online) The arrows in panel a) $(\textcolor{blue}{\longrightarrow},\textcolor{red}{\longrightarrow},\textcolor{green}{\longrightarrow})$ correspond to $(\theta_1,\theta_2,\theta_3)$. Panels (b) and (c) show examples of phase configurations for the two $\groupZ$ symmetry classes of the ground states, shown on a $2\times 2$ lattice of selected points of a planar slice of the system. Here $g_{12} > g_{23} > g_{13} > 0$. The spatial contribution to the energy is minimized by making the spatial gradient zero (hence breaking the global $\groupU$ symmetry). Then there are two classes of phase configurations, one with chirality +1 and one with chirality -1, minimizing the energy associated with the interband interaction. The chirality is defined as $+1$ if the phases (modulo $2\cpi$) are cyclically ordered $\theta_1<\theta_2<\theta_3$, and $-1$ if not.}  
\label{fig:gr_state_Z2}
\end{figure}

When multiple (three or more) complex order parameters are needed to describe the condensate of the superfluid or superconductor,  an additional $\groupZ$ (\enquote{time reversal}) symmetry may be needed for describing the system.\cite{PhysRevB.88.220511,PhysRevB.89.104509} Such a situation is expected to occur in the iron-pnictides in some parameter regime,\cite{zlatko,maiti} but will also occur in other systems involving more than two superconducting order-parameter components  where several superconducting bands cross the Fermi level, interacting with each other through Josephson couplings.\cite{nagaosa,stanev,johan3,PhysRevB.89.104509,PhysRevB.88.220511} For repulsive Josephson couplings, the resulting frustration leads to two classes of (mirrored) $\groupU$ symmetric ground states. Hence, the system features an overall $\groupUZ$ symmetry. This is illustrated in \cref{fig:gr_state_Z2}. For details, see Refs. \onlinecite{PhysRevB.88.220511,PhysRevB.89.104509}.

In Ref.~\onlinecite{PhysRevB.88.220511} it was shown that in a multiband $\groupUZ$ superconductor only the $\groupU$ sector, and not the $\groupZ$ sector, couples to a gauge field. Hence, if we induce vortices in such a superconductor by an external field, the behavior of the $\groupZ$ sector is expected to be largely unaffected. Thus, by applying an external field to a $\groupUZ$ superconductor, one should be able to control the $\groupU$ sector independently of the $\groupZ$ sector, an effect which should be experimentally detectable. 

Of special interest is the study of the $\groupU$ symmetric, but $\groupZ$ broken metallic phase predicted to be present in the multiband superconductors for a range of parameters.\cite{PhysRevB.89.104509,PhysRevB.88.220511} We show that by tuning an external magnetic field, it is possible to extend the region of the $\groupZ$ broken metallic phase in the phase diagram.

\section{Models}
In this work, we consider two versions of a 3D minimal $n$-component model in the London limit of the Ginzburg-Landau model of a multiband superconductor, displaying $\groupUZ$ symmetry. We focus on the simplest non-trivial case of three components. Both versions of the model are described in greater detail in Ref.~\onlinecite{PhysRevB.89.104509}, see also reference therein. In particular, it has been shown \cite{weston} that the inclusion of more than three superconducting bands crossing the Fermi surface will, apart from states with measure zero in parameter space, yield the same physics as in the three-band case.  We include a non-fluctuating $\groupU$ gauge field with a tunable value in the description, which in turn will lead to induced vortices. Neglecting the fluctuations in the amplitudes of the order-parameter components and the $\groupU$ gauge-field is consistent with the fact that the pnictide superconductors are in the extreme type-II regime. \cite{Nature3741995,PhysRevB.59.6449,PhysRevB.60.15307} The use of the London limit therefore rests on solid ground in this case.

\subsection{Full model}
The model on the $L^3$ lattice (with periodic boundary conditions) is given by
\begin{multline}
 H = -\sum_{\langle i,j\rangle,\alpha } a_{\alpha}\cos (\theta_{\alpha,i} - \theta_{\alpha,j} - A_{ij}) \\
 + \sum_{i,\alpha' > \alpha} g_{\alpha \alpha'} \cos (\theta_{\alpha,i} - \theta_{\alpha',i}),
 \label{eq:lattice_full_H}
\end{multline}
where the gauge field is chosen to be
\begin{equation}
 \v{A}(\v{r}) = (2\cpi yf, 0, 0).
\end{equation}
$i$ and $j$ are lattice site indices and $\langle i,j\rangle$ denote nearest neighbor sites. $\alpha,\alpha' \in \set{1,2,\ldots,n}$ are component labels. $f$ is the vortex filling fraction, which is a direct measure of the rotation of the system. Moreover, $a,g > 0$ are parameters determining the condensate density and intercomponent Josephson interaction, respectively. For convenience $a_1$ is set to $a_1 = 1$. Note that we have rescaled the gauge field $\v A$ with the electric charge $e$, $\v A \gets e \v A$.

\subsection{Reduced ($K_1 K_2$) model}
Previous works \cite{PhysRevB.89.104509} have shown that the interband fluctuations of the phases of the \enquote{full} model, \cref{eq:lattice_full_H}, are not of qualitative importance when mapping out the phase diagram. These fluctuations may be suppressed by letting $g_{\alpha\alpha'} \to \infty$ while keeping the ratios $g_{\alpha\alpha'}/g_{\alpha''\alpha'''}$ finite, locking the phase \enquote{stars} to one of their two ground state configurations, see \cref{fig:3C_ground_state_1,fig:3C_ground_state_2}. The advantage of doing so is twofold. First, the $\groupUZ$ structure of the system is brought out clearly. Secondly, the computational cost of simulations is significantly reduced,\footnote{The reduced model is less computationally demanding than the full model for several reasons, the most important one being the reduction of degrees of freedom per lattice site.} meaning that larger systems and better statistics are obtainable.



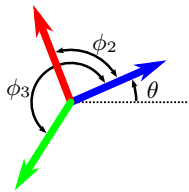
\begin{figure}
  \def\svgwidth{0.4\columnwidth}
  \begingroup%
    \makeatletter%
    \providecommand\color[2][]{%
      \errmessage{(Inkscape) Color is used for the text in Inkscape, but the package 'color.sty' is not loaded}%
      \renewcommand\color[2][]{}%
    }%
    \providecommand\transparent[1]{%
      \errmessage{(Inkscape) Transparency is used (non-zero) for the text in Inkscape, but the package 'transparent.sty' is not loaded}%
      \renewcommand\transparent[1]{}%
    }%
    \providecommand\rotatebox[2]{#2}%
    \ifx\svgwidth\undefined%
      \setlength{\unitlength}{114.7145525bp}%
      \ifx\svgscale\undefined%
	\relax%
      \else%
	\setlength{\unitlength}{\unitlength * \real{\svgscale}}%
      \fi%
    \else%
      \setlength{\unitlength}{\svgwidth}%
    \fi%
    \global\let\svgwidth\undefined%
    \global\let\svgscale\undefined%
    \makeatother%
    \begin{picture}(1,0.70174618)%
      \put(0,0){\includegraphics[width=\unitlength]{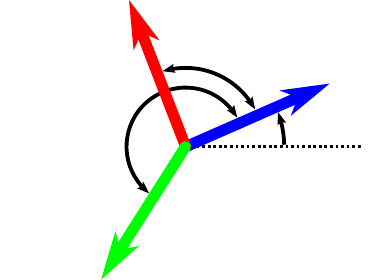}}%
      \put(0.76700319,0.35609009){\color[rgb]{0,0,0}\makebox(0,0)[lb]{\smash{$\theta$}}}%
      \put(0.5569646,0.54141848){\color[rgb]{0,0,0}\makebox(0,0)[lb]{\smash{$\phi_{2}$}}}%
      \put(0.32367835,0.3684454){\color[rgb]{0,0,0}\makebox(0,0)[rb]{\smash{$\phi_{3}$}}}%
    \end{picture}%
  \endgroup%
  \caption{(Colors online) One of the two $\groupZ$ phase configurations in the $g_{\alpha\alpha'}\to \infty$ limit when $n=3$. $\phi_\alpha$ is the phase difference between the first and the $\alpha$'th component, a constant.}
\label{fig:K1K2_derivation}
\end{figure}

The Hamiltonian may now be written as\cite{PhysRevB.89.104509}
\begin{multline}
 H = -\sum_{\langle i,j\rangle} (1 + K_1 \sigma_i\sigma_j)\cos (\theta_{i} - \theta_{j} - A_{ij}) \\
 -\sum_{\langle i,j\rangle} K_2(\sigma_i - \sigma_j)\sin (\theta_{i} - \theta_{j} - A_{ij})
 \label{eq:H_reduced}
\end{multline}
Here $\sigma_j$ is a statistically fluctuating Ising-variable on each lattice site, denoting the chirality of the phase-star, while $\theta_j$ is a statistically varying $\groupU$-variable denoting the overall orientation of the phase-star, see \cref{fig:K1K2_derivation} as well as \cref{fig:3C_ground_state_1,fig:3C_ground_state_2}.
$K_1$ and $K_2$ are parameters given by
\begin{align}
 K_1 \equiv& \frac{\sum_{\alpha>1} a_\alpha\bigl[1-\cos(2\phi_{\alpha})\bigr]}{2 + \sum_{\alpha>1} a_\alpha \bigl[1+\cos(2\phi_{\alpha})\bigr]}\\
 K_2 \equiv& \frac{\sum_{\alpha>1} a_\alpha \sin(2\phi_{\alpha})}{2 + \sum_{\alpha>1} a_\alpha \bigl[1+\cos(2\phi_{\alpha})\bigr]},
\label{eq:K1K2}
\end{align}
where $\phi_{\alpha}$ is the phase difference between the the first and the $\alpha$'th component in the ground state phase star, as illustrated in \cref{fig:K1K2_derivation}.

It can be shown that $K_1$ and $K_2$ are restricted to the ellipsis given by 
\begin{equation}
 \left[\frac{2}{n-1}K_1 - 1\right]^2 + \left[\frac{2\sqrt{n}}{n-1}K_2\right]^2 \leq 1.
 \label{eq:domain}
\end{equation}
  
An equivalent formulation to \cref{eq:H_reduced} reads (see \cref{app:alternative_reduced})
\begin{equation}
 H = -\sum_{\langle i,j \rangle} (1 + J \sigma_i\sigma_j)\cos(\theta_i - \theta_j - A_{ij} - \alpha(\sigma_i,\sigma_j))
 \label{eq:H_reduced_alternative}
\end{equation}
where
\begin{equation}
 \alpha(\sigma_i,\sigma_j) \equiv \begin{cases}
      0 & \sigma_i = \sigma_j \\
      \pm \arctan\left[\frac{2K_2}{1 - K_1 }\right] &\sigma_i = -\sigma_j = \pm 1
     \end{cases}
\end{equation}
and
\begin{equation}
 J = \frac{W}{1 + \sqrt{1 - W^2}} \in [0,K_1],
 \label{eq:J}
\end{equation}
where
\begin{equation}
 W \equiv \frac{2(K_1 - {K_2}^2)}{1 + {K_1}^2 + 2{K_2}^2}.
 \label{eq:W}
\end{equation}

\Cref{eq:H_reduced_alternative} reveals an interesting feature of the model. $K_2 \neq 0$, i.e. when the system features a deviation from a symmetric ground state phase star (i.e. $\phi_{12} \neq \phi_{13} \neq \phi_{23}$ in the $n=3$ case), leads to the addition of a fluctuating quantity coupling minimally to the phase-difference  on a link, $\theta_i-\theta_j$. It formally has the appearance of a fluctuating \emph{discrete} \enquote{gauge field}, $\alpha$, in a Ising-XY model. It should be kept in mind, however, that $\alpha(\sigma_i,\sigma_j)$ is only a \enquote{semi-independent} degree of freedom since it couples to the prefactor through the $J\sigma_i\sigma_j$ term.

\section{Observables}

In the full model, as well as in the reduced one, the $\groupZ$ sector is monitored by the (global) \enquote{magnetization} defined as
\begin{equation}
 m \equiv N\inv\sum_i \sigma_i
\end{equation}

We use the Binder cumulant,\cite{Binder_1981,Sandvik_2010}
\begin{equation}
 U_2 \equiv \frac{1}{2}\left(3-\frac{\avg{m^4}}{\avg{m^2}^2}\right),
\end{equation}
to detect phase transitions. The Binder cumulant displays a non-analytical jump at the phase transition in the thermodynamical limit, and has the useful property of being only mildly affected by finite size effects.

For the reduced model, we use the helicity modulus along the $z$-axis, the direction of the external field, to probe the structural order of the vortex system. Furthermore, to make sure that there is no pinning of the vortices to the underlying numerical lattice, we monitor the helicity modulus in the $x$ and $y$ directions as well. These should be zero for all temperatures of interest if such numerical artifacts are to be avoided.

In the full model, the helicity modulus is no longer well defined if one wants to consider the formation of a vortex lattice in each of the individual components. We choose therefore instead to use the value of the planar structure function of the vortices at the first Bragg peak to monitor the vortex lattice as the temperature is varied. In the liquid phase this will be a small number (approaching zero in the thermodynamical limit), while in the ordered phase this number will be finite. The structure function for a given momentum $\v{k}_{\perp}$ in the plane perpendicular to the direction of the external field, the $xy$ plane, is given by
\begin{equation}
 S^{\alpha}(\v{k}_{\perp}) = \frac{1}{(fL^3)^2}\biggl\langle\Bigl|\sum_{\v r} n_{z}^{\alpha}(\v r) \e{\i \v{k}_{\perp}\cdot \v{r}_{\perp}} \Bigr|^2 \biggr\rangle
\end{equation}
$\v{r}_{\perp}$ is the projection of the position vector $\v{r}$ onto the $xy$ plane. $\v{n}^\alpha(\v{r})$ is the vorticity vector (which can be $0,\pm 1$ in each spatial component) of component $\alpha$ of the field in point $\v{r}$,
\begin{equation}
 \v{n}^{\alpha}(\v{r}) = \frac{1}{2\cpi}[\nabla\times(\nabla \theta_{\alpha} - e\v{A})]
\end{equation}

We also monitor the specific heat,
\begin{equation}
 c \equiv N\inv C = N\inv \beta^2 \big\langle(H-\avg{H})^2\big\rangle.
\end{equation}

\section{Simulations and Results}
Due to the frustration effects inherent in the models, there appears to be no efficient nonlocal (cluster) algorithm for simulating them. Hence,  a local update Monte Carlo scheme, the \enquote{Fast Linear Algorithm} (FLA) of Ref. \onlinecite{Loison_et_al_2004}, was used. It proved to be a significant improvement over the standard Metropolis-Hastings sampling, and appears to be the most efficient canonical algorithm available for the models investigated in this paper. However, for technical reasons the use of FLA meant that we were prevented from simulating the case $K_1 = 1$ of the reduced model, where the effect of the intraband frustration is strongest in a \emph{three} component reduced model. Therefore, in the simulations, the parameter-value $K_1=0.99$ was chosen as a reasonable compromise between proximity to $K_1=1$ and numerical stability.

In order to take advantage of the computational resources available, grid parallellization was implemented. Ferrenberg-Swendsen multi-histogram reweighting \cite{Ferrenberg_PRL_1989} was used to improve our numerical data. Pseudorandom numbers were generated by the Mersenne--Twister algorithm~\cite{Matsumoto_1998_ACM_TMCS}.

The reduced model is significantly less computational demanding than the full model, and most of the simulations were performed on the former. To demonstrate the equivalence of the two models, we first show that the full model gives equivalent results to the reduced model for a representative choice of parameters. (See also Ref.~\onlinecite{PhysRevB.89.104509}.) 

Moreover, we establish the main point conjectured earlier, namely that an external field separates the $\groupZ$ transition and the $\groupU$ lattice melting, with separation increasing with field strength since the external gauge-field couples to the $\groupU$-sector, but not the $\groupZ$- sector. Thus, as magnetic field is increased, we observe a reversal of the order of the $\groupU$ and $\groupZ$ transitions as a function of temperature. 

\begin{figure}[ht]
 \includegraphics{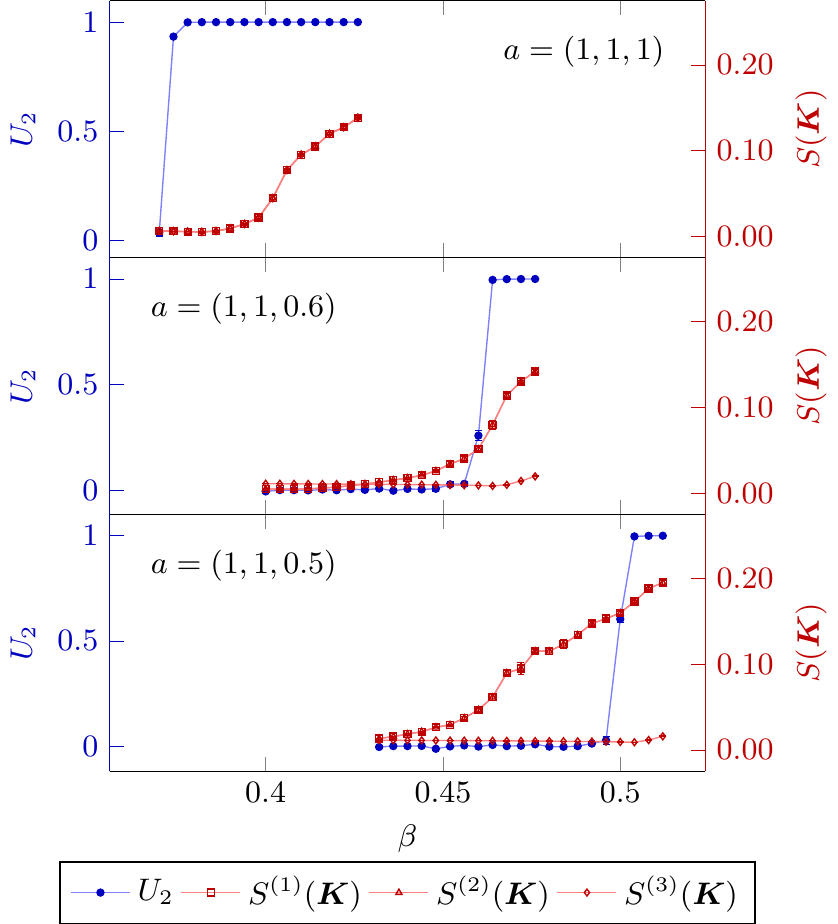}
\caption{Simulation results for the full model with $a$ as indicated, and $g = (5,5,5)$, $f=\tfrac{1}{32}$, and $L=64$. 
The $\groupZ$ sector is monitored by the Binder cumulant for the $\groupZ$ magnetization, $U_{2}$ (blue). The vortex lattice ordering (in each component) is monitored by the value of the structure function at the first Bragg peak, $S(\v K)$ (red). Note how the order of the phase transitions changes as the anisotropy of the system $a_3/a_{1,2}$ is varied. In the top panel, the system features a normal metallic state (vortex liquid), with broken $\groupZ$ symmetry being restored at $\beta \approx 0.37$. In the bottom panel, the $\groupZ$ symmetry is restored inside the superconducting (vortex lattice) phase. In the middle panel, the transitions occur roughly simultaneously. (Colors online.)\label{fig:full}}
\end{figure}

\Cref{fig:full} shows simulation results from the full model, \cref{eq:lattice_full_H}, for various  different  choices $a_3$ of the model, with fixed $a_1=a_2=1$. This variation effectively leads to a variation in the angles $\phi_\alpha$  describing the relative orientations of the various phases of the components of the order parameter in the ground state, and hence to a variation in the energy of the $\groupZ$ domain walls of the system. This in turn will lead to a variation in the critical temperature of the $\groupZ$ phase transition responsible for restoring time-reversal symmetry. The rotation of the system is fixed at a filling fraction $f=\tfrac{1}{32}$.  

The top panel of \cref{fig:full} corresponds to the fully symmetric case where all $a_{\alpha}$ and $g_{\alpha,\alpha'}$ are equal in \cref{eq:lattice_full_H}, in turn corresponding to the case $K_2=0$ in \cref{eq:K1K2}. Reducing $a_3$ in the following panels shifts the $\groupZ$ phase transition downwards in temperature {as the domain wall energy decreases}. The transition temperature in the $\groupU$-sector, in this case the vortex-lattice melting transition, is little affected by the reduction in $a_3$, since the vortex-lattice melting temperature is largely determined from the phase-stiffness of the overall phase-star, and not the relative-fluctuations of the internal phases of the multi-component order parameter. The former stiffness is dominated by the largest phase-stiffnesses of the individual phases, see for instance Eq.~(2) of Ref. \onlinecite{PhysRevB.88.220511}. Thus, the $\groupZ$ transition temperature is eventually lowered through the $\groupU$ transition temperature. This reversal of the phase transition of the $\groupZ$ and $\groupU$ sectors means that the system transitions from one featuring a superconducting state with broken time-reversal symmetry and a time-reversal symmetric metal, to one with a time-reversal symmetric superconducting state and a metallic state with a spontaneously broken time-reversal symmetry. Below, we return to the experimental probes of the $\groupZ$ phase transition inside the superconducting or metallic states. 

\begin{figure}[ht]
 \includegraphics{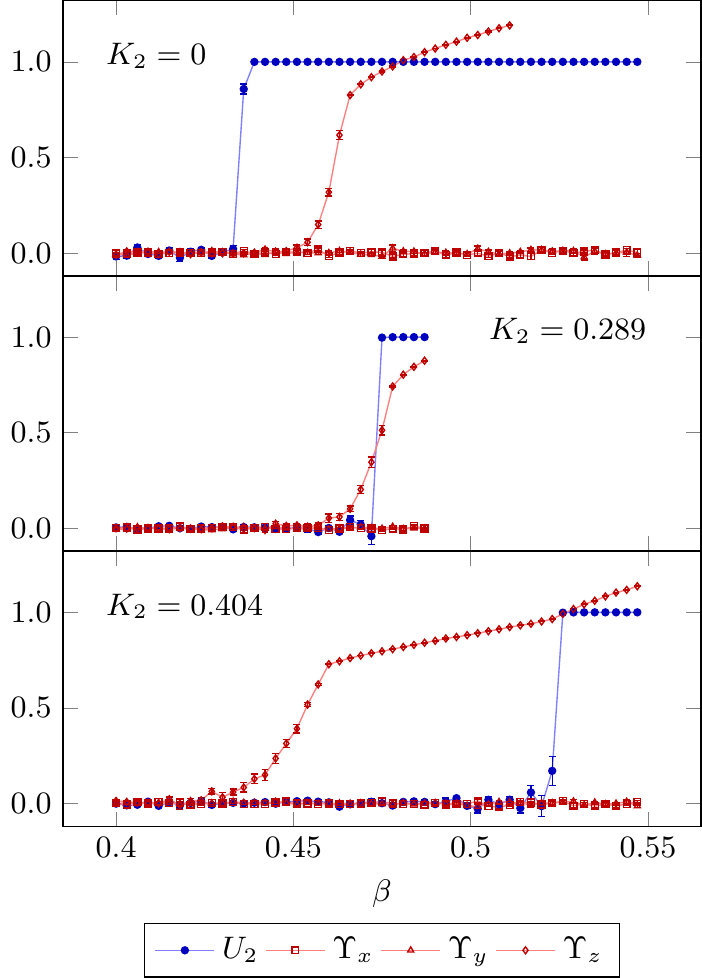}
\caption{Simulation results for the reduced model with $K_1 = 0.99$, $f=\tfrac{1}{32}$, $L=128$ and $K_2 = 0, \num{0.058},\num{0.173}, \num{0.289}, \num{0.404}$. The choice of the parameter $K_1$ is explained in the text. The $\groupZ$ sector is monitored by the Binder cummulant for the $\groupZ$ magnetization, $U_{2}$ (blue). The vortex lattice ordering is monitored by the helicity modulus, $\Upsilon$, (red) in the various directions, where $z$ is parallel to the external field. (Colors online.)\label{fig:var_K2}}
\end{figure}

The results of \cref{fig:full} should be compared with the results for the reduced model, \cref{fig:var_K2}. For $K_2=0$, corresponding to the results shown in the upper panel of \cref{fig:full}, {{the same result is found for smaller $K_2$ values, i.e}} the $\groupZ$ transition is found at higher temperature than the $\groupU$ due to the relatively large energy of the $\groupZ$ domain-walls. As $K_2$ increases, the relative energy associated with a $\groupZ$ domain wall decreases, eventually resulting in a reversal of the order of the $\groupZ$ and $\groupU$ transitions. These effects are thus essentially the same in the full and reduced models. 

\begin{figure}[ht]
 \includegraphics{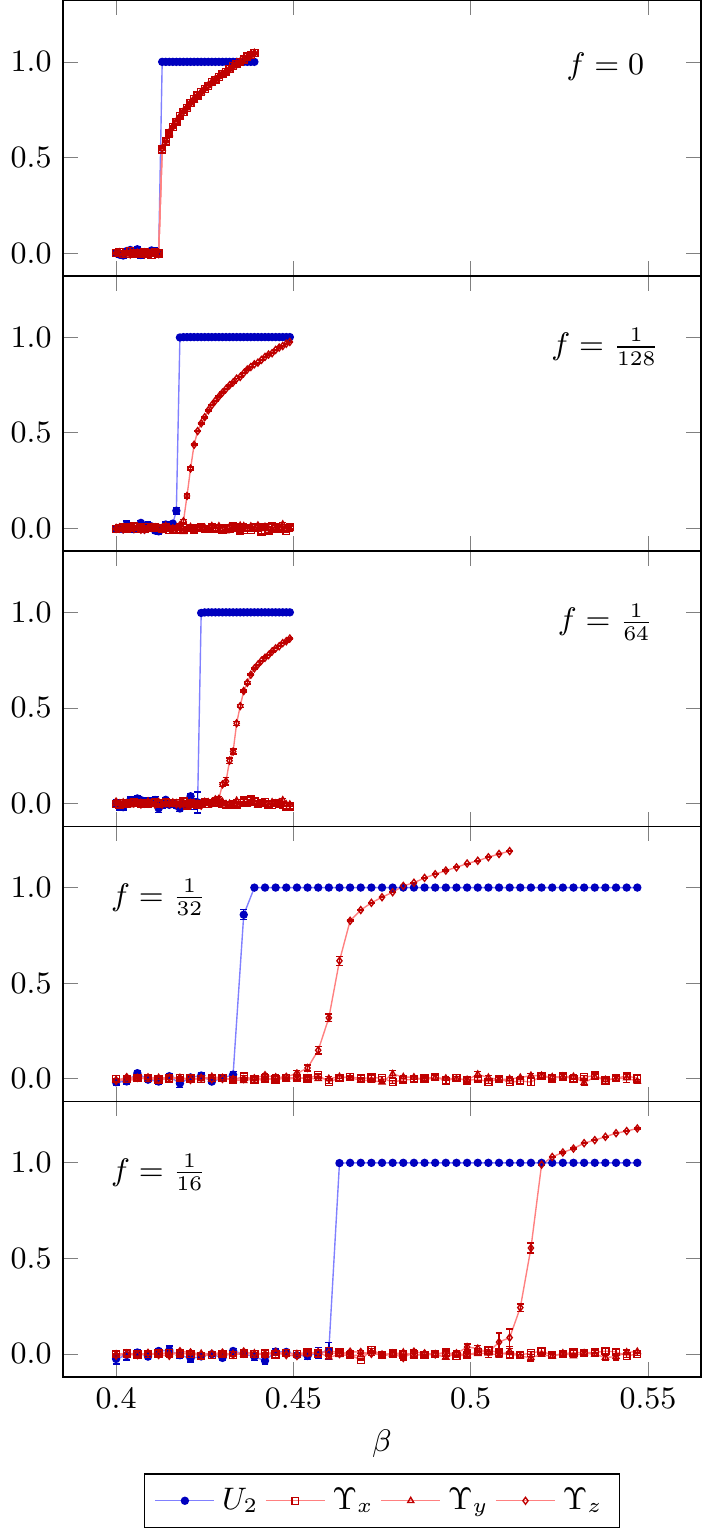}
\caption{Simulation results for the reduced model with $K_1 = 0.99$, $K_2 = 0.0$, $L=128$ and $f = 0, \tfrac{1}{128}, \tfrac{1}{64}, \tfrac{1}{32}, \tfrac{1}{16}$. The $\groupZ$ sector is monitored by the Binder cumulant for the $\groupZ$ magnetization, $U_{2}$ (blue). The vortex lattice ordering is monitored by the helicity modulus, $\Upsilon$, (red) in the various directions, where $z$ is parallel to the external field. In zero external field ($f=0$) the system is isotropic. (Colors online.)\label{fig:var_f}}
\end{figure}

We next consider the effect of varying the rotation $f$ at otherwise fixed parameters. For this, we limit the discussion to the reduced model, \cref{eq:H_reduced}. \Cref{fig:var_f} show how, as conjectured, the separation of the $\groupZ$ and $\groupU$ transitions increase with an increasing external field strength. To work with a manageable parameter space, we limit the study to $K_2=0$ since this suffices to illustrate our main point, namely the separation of two otherwise simultaneous zero-field phase transitions when the field strength is increased. For the special case $f=0$, the $\groupZ$ and $\groupU$ transitions occur simultaneously via a preemptive first-order mechanism, and there is never a chiral metallic state in the absence of a fluctuating gauge field \cite{PhysRevB.89.104509}.  As the field strength is increased, the transitions separate, with the $\groupU$ transition being strongly suppressed to lower temperature while the $\groupZ$ transition remains only weakly affected. This follows from the fact that it is only the $\groupU$-sector of the theory which couples to the (non-fluctuating) gauge field, while the $\groupZ$-sector does not. Hence, upon increasing the (non-fluctuating) gauge-field and hence the filling fraction of the system, the vortex-lattice melting transition of the $\groupU$-sector is suppressed in the usual manner, while the $\groupZ$-sector is largely unaffected.  A reversal of the order of the phase-transitions as the temperature is varied, is thus possible. An increase of the magnetic field beyond the vortex-lattice melting transition brings about a  resistive state with spontaneously broken $\groupZ$-symmetry, a chiral metallic state. 

Note that the temperature dependence of the structure function in Fig. \ref{fig:full} and the helicity moduli in \cref{fig:var_K2,fig:var_f}  typically is not of the form one expects in a first-order vortex lattice transition, with a jump in the helicity modulus at the melting transition, and which has been found in the single-component case \cite{PhysRevLett.79.3498,PhysRevB.60.15307}. This point requires further investigation, but is beyond the scope of the present paper, where the main point is not to investigate the details of the melting transition, but to demonstrate that a magnetic field may be utilized to clearly bring out the unusual metallic state with a spontaneously broken time-reversal symmetry. 

\begin{figure}[ht]
 \includegraphics{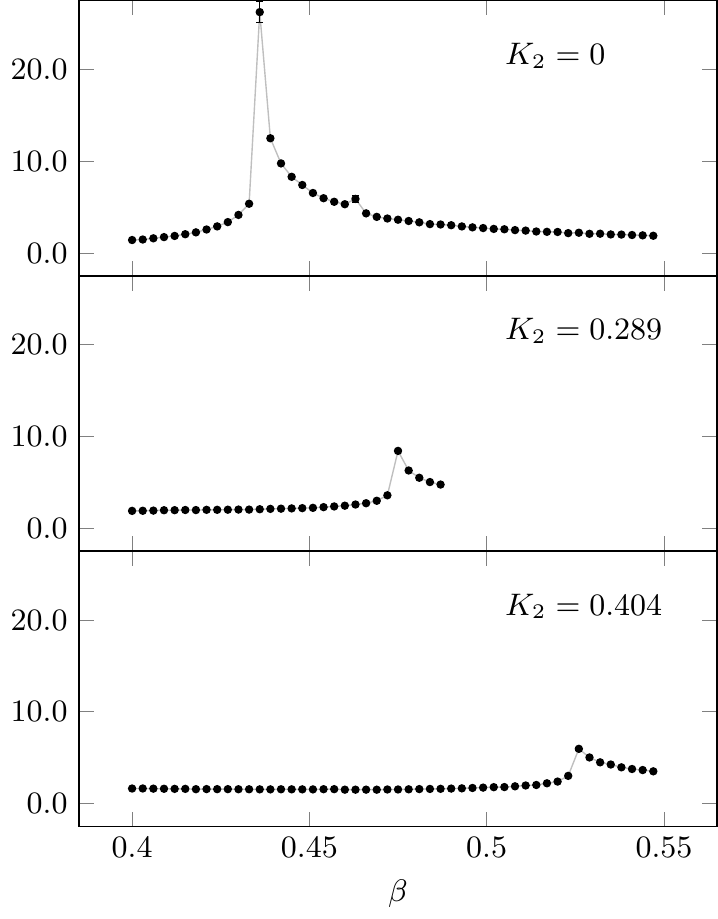}
\caption{Specific heat capacity of the reduced model with $K_1 = 0.99$, $f=\tfrac{1}{32}$, $L=128$ and $K_2 = 0, \num{0.289}, \num{0.404}$. Associated with \cref{fig:var_K2}. \label{fig:c_var_K2}}
\end{figure}

\begin{figure}[ht]
 \includegraphics{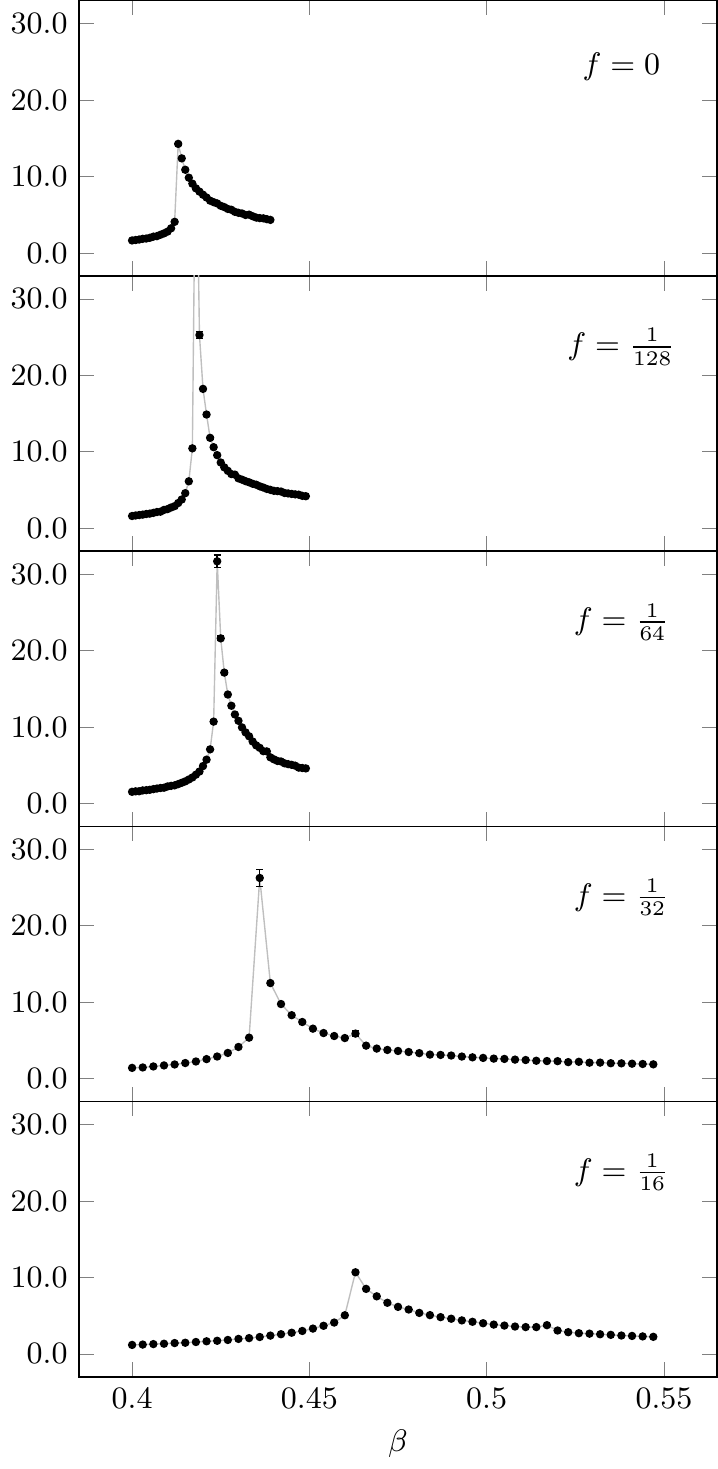}
\caption{Specific heat of the reduced model with $K_1 = 0.99$, $K_2 = 0.0$, $L=128$ and $f = 0, \tfrac{1}{128}, \tfrac{1}{64}, \tfrac{1}{32}, \tfrac{1}{16}$. These results correspond to those shown in  \cref{fig:var_f}. Note that the $\groupZ$-anomaly is considerably more prominent than the sharp peak associated with the 
melting of the vortex lattice. \label{fig:c_var_f}}
\end{figure}

\Cref{fig:c_var_K2,fig:c_var_f} show the specific heat corresponding to the results of  \cref{fig:var_K2,fig:var_f}. The main point to be made in connection with \cref{fig:c_var_K2,fig:c_var_f} is that the Ising-type anomaly in the vortex-liquid phase, associated with restoring the spontaneously broken time-reversal symmetry, is considerably more pronounced than the small anomaly associated with the 
vortex lattice melting. These two phase-transitions essentially involve the same degrees of freedom, ultimately connected to the phases $\theta_{\alpha,j}$ of the superconducting order-parameter components. Hence, to the extent that the specific heat anomaly associated with the vortex-lattice melting is observable, the Ising-anomaly inside the vortex-liquid phase, associated with restoring the broken $\groupZ$-symmetry, should be readily observable in specific-heat measurements on Fe-pnictides. Moreover, an equally prominent $\groupZ$-anomaly in the specific heat should be observable inside the vortex-lattice state, for small enough magnetic fields. 

\section{Summary and conclusions}
We have studied two models describing $\groupUZ$ multiband superconductors in the London limit, subject to an external field. We have focused on the three-component case. The external field induces vortices in the condensate, leading to an increased separation of, and indeed reversal of, the $\groupZ$ and $\groupU$ phase transitions as the temperature is varied. This brings out clearly the domain of a metallic (vortex liquid state) with an additional spontaneously broken time-reversal symmetry on top of the explicitly broken time-reversal symmetry from the external field. The effect increases with increasing field. Inside the vortex-liquid phase there should be an anomaly in the specific heat, and this anomaly should be in the 3D Ising universality class. The same degrees of freedom are involved in disordering the vortex lattice as are involved in disordering the chirally ordered state. The numeric results show that both anomalies are observable, but the $\groupZ$-anomaly is considerably easier to see, see \cref{fig:c_var_K2,fig:c_var_f}. Hence, we expect that this anomaly associated with restoring the $\groupZ$ chiral order to be readily observable in experiments. Moreover, the same should be the case for the $\groupZ$-anomaly in the specific heat inside the vortex-lattice for small enough magnetic fields. Finally, we note that the predictions of anomalies in the specific, obtained in the London-limit of the Ginzburg-Landau theory of a multi-band superconductor, 
should be robust to inclusion of amplitude fluctuations in the order-parameter components. Such fluctuations are non-critical, but will nonetheless tend to enhance the specific heat-anomalies, albeit analytically as a function of temperature.

T.A.B. thanks NTNU for financial support, and the Norwegian consortium for high-performance computing (NOTUR) for computer time. A.S. was supported by the Research Council of Norway, through Grants 205591/V20 and 216700/F20. AS thanks the Aspen Center for Physics (NSF Grant No 1066293) for hospitality during the initial stages of this work. 

\appendix
\section{Derivation of alternative reduced model \label{app:alternative_reduced}}
We derive \cref{eq:H_reduced_alternative} from \cref{eq:H_reduced}.

The identity
\begin{equation}
 A\cos x + B\sin x = \sqrt{A^2 + B^2}\cos\left[x - \arctan\left(\tfrac{B}{A}\right)\right],
 \label{eq:trig_linear_comb}
\end{equation}
together with $\sigma_i^2 = 1$,  implies that the contribution from a lattice link to the Hamiltonian, \cref{eq:H_reduced}, can be written on the form 
\begin{align}
 H_{ij} = {}& - (1 + K_1 \sigma_i\sigma_j)\cos (\theta_{i} - \theta_{j} - A_{ij}) \nonumber\\
 & - K_2(\sigma_i - \sigma_j)\sin (\theta_{i} - \theta_{j} - A_{ij}) \nonumber\\
 = {}& -(p + q\sigma_i\sigma_j)\cos (\theta_{i} - \theta_{j} - A_{ij} - \alpha(\sigma_i,\sigma_j)).
 \label{eq:Hij}
\end{align}
$p$,$q$, and $\alpha$ are functions of $K_1$ and $K_2$, to be determined. 

Comparing with \cref{eq:trig_linear_comb}, it is  seen that $\alpha$ is given by
\begin{align}
 \alpha &= \arctan\left[\frac{K_2(\sigma_i - \sigma_j)}{1 + K_1 \sigma_i\sigma_j}\right] \notag\\
  &= \begin{cases}
      0 & \sigma_i = \sigma_j \\
      \pm \arctan\left[\frac{2K_2}{1 - K_1 }\right] &\sigma_i = -\sigma_j = \pm 1
     \end{cases}
\end{align}

Similarly, $p$ and $q$ are determined by
\begin{equation}
 \sqrt{(1 + K_1 \sigma_i\sigma_j)^2 + {K_2}^2(\sigma_i + \sigma_j)^2} = p + q \sigma_i \sigma_j,
\end{equation}
or, by squaring both sides,
\begin{equation}
 1 + {K_1}^2 + 2{K_2}^2 + 2(K_1 - {K_2}^2)\sigma_i\sigma_j = p^2 + q^2 + 2pq\sigma_i\sigma_j.
\end{equation}
Comparing the two sides, we see that
\begin{align}
 p^2 + q^2 &= 1 + {K_1}^2 + 2{K_2}^2 \equiv U, \\
 pq &= K_1 - {K_2}^2 \equiv V.
\end{align}
Combining these two equations,  yields the quadratic equation in $p^2$
\begin{equation}
 p^4 - Up^2 + V^2 = 0,
\end{equation}
with solutions  
\begin{equation}
 p = \pm\sqrt{\tfrac{1}{2}\left(U \pm \sqrt{U^2 - 4V^2}\right)}.
\end{equation}
When $K_2 = 0$, $p + q\sigma_i\sigma_j$ should reduce to $1 + K_1\sigma_i\sigma_j$. Hence, the relevant solution is 
\begin{align}
 p &= \sqrt{\tfrac{1}{2}\left(U + \sqrt{U^2 - 4V^2}\right)}, \\
 q &= \frac{V}{p}.
\end{align}
Rescaling the Hamiltonian by $1/p$ simplifies \cref{eq:Hij} to
\begin{equation}
 H_{ij} = -\left(1 + J \sigma_i\sigma_j\right)\cos (\theta_{i} - \theta_{j} - A_{ij} - \alpha(\sigma_i,\sigma_j)),
\end{equation}
where $J$ is given by \cref{eq:J,eq:W}. Thus, since $H = \sum_{\langle i,j \rangle} H_{ij}$, we have derived \cref{eq:H_reduced_alternative}.

For $n\geq 3$, we have $W \in [0,2K_1/(1 + {K_1}^2) < 1]$, since $K_1 \in [0,n-1]$ and $K_1 \geq {K_2}^2$. $J(W)$ increases monotonically with $W$ for $W\in[0,1)$, so
\begin{equation}
 J \in [0,K_1].
\end{equation}

  \bibliography{references}

%
%
%
%

\end{document}